\documentclass[sn-mathphys-num]{sn-jnl}

\newcommand{\araa}{Annu. Rev. Astron. Astrophys.}   
\newcommand{\aj}{Astron. J.}   
\newcommand{\apj}{Astrophys. J.}   
\newcommand{\apjl}{Astrophys. J. Lett.}   
\newcommand{\aap}{Astron. Astrophys.}   
\newcommand{\bain}{Bulletin of the Astronomical Institutes of the Netherlands} %
\newcommand{\mnras}{Mon. Not. R. Astron. Soc.}   
\newcommand{\nat}{Nature} 
\newcommand{\pasp}{Publ. Astron. Soc. Pac.}   
\newcommand{\ssr}{Space Sci. Rev.}   

\usepackage{graphicx}%
\usepackage{multirow}%
\usepackage{amsmath,amssymb,amsfonts}%
\usepackage{amsthm}%
\usepackage{mathrsfs}%
\usepackage[title]{appendix}%
\usepackage{xcolor}%
\usepackage{textcomp}%
\usepackage{manyfoot}%
\usepackage{booktabs}%
\usepackage{algorithm}%
\usepackage{algorithmicx}%
\usepackage{algpseudocode}%
\usepackage{listings}%
\usepackage{fullpage}
\usepackage{siunitx}
\usepackage{bm}
\newcommand{\molh}{H$_2$}
\newcommand{\mic}{$\mu$m}
\newcommand{\kms}{km\,s$^{-1}$}

\newcommand{\msun}{$M_\odot$}
\newcommand{\mstar}{$M_\star$}
\newcommand{\mhi}{$M_\mathrm{H\textsc{i}}$}
\newcommand{\hi}{H\,\textsc{i}}
\newcommand{\hii}{H\,\textsc{ii}}
\newcommand{\mmolh}{$M_{\mathrm{H}_2}$}
\newcommand{\mwarm}{$M_{\mathrm{H}_2}^\mathrm{warm}$}

\newcommand{\lco}{$L_\mathrm{CO}$}
\newcommand{\aco}{$\alpha_\mathrm{CO}$}

\begin{document}

\title{Molecular Hydrogen in the Extremely Metal-Poor, Star-Forming Galaxy Leo~P}

\author[1,2,3]{\fnm{O. Grace} \sur{Telford}}
\author[4]{\fnm{Karin M.} \sur{Sandstrom}}
\author[3,5]{\fnm{Kristen B. W.} \sur{McQuinn}}
\author[6]{\fnm{Simon C. O.} \sur{Glover}}
\author[5]{\fnm{Elizabeth J.} \sur{Tarantino}}
\author[7,8]{\fnm{Alberto D.} \sur{Bolatto}}
\author[4,5]{\fnm{Ryan J.} \sur{Rickards Vaught}}

\affil[1]{Department of Astrophysical Sciences, Princeton University, 4 Ivy Lane, Princeton, NJ 08544, USA; {\color{blue}grace.telford@princeton.edu}}
\affil[2]{The Observatories of the Carnegie Institution for Science, 813 Santa Barbara Street, Pasadena, CA 91101, USA}
\affil[3]{Department of Physics and Astronomy, Rutgers University, 136 Frelinghuysen Road, Piscataway, NJ 08854, USA}
\affil[4]{Department of Astronomy \& Astrophysics, University of California, San Diego, La Jolla, CA 92093, USA}
\affil[5]{Space Telescope Science Institute, 3700 San Martin Drive, Baltimore, MD 21218, USA}
\affil[6]{Universit\"{a}t Heidelberg, Zentrum f\"{u}r Astronomie, Institut f\"{u}r Theoretische Astrophysik, Albert-Ueberle-Str. 2, D-69120 Heidelberg, Germany}
\affil[7]{Department of Astronomy, University of Maryland, College Park, MD 20742, USA}
\affil[8]{Joint Space-Science Institute, University of Maryland, College Park, MD 20742, USA}

\maketitle 
\vspace{-40pt}
\textbf{The James Webb Space Telescope (JWST) has revealed unexpectedly rapid galaxy assembly in the early universe, in tension with models of star and galaxy formation \cite{labbe23, boylan-kolchin23, carniani24}. 
In the gas conditions typical of early galaxies, particularly their low abundances of heavy elements (metals) and dust, the star-formation process is poorly understood. 
Some models predict that stars form in atomic gas at low metallicity \cite{glover12, krumholz12}, in contrast to forming in molecular gas as observed in higher-metallicity galaxies \cite{schinnerer24}. 
To understand the very high star-formation rates at early epochs, it is necessary to determine whether molecular gas formation represents a bottleneck to star formation, or if it is plentiful even at extremely low metallicity.
Despite repeated searches \citep{warren15}, star-forming molecular gas has not yet been observed in any galaxy below 7\% of the Solar metallicity \cite{Shi2020}, leaving the question of how stars form at lower metallicities unresolved. 
Here, we report the detection of rotationally excited emission from molecular hydrogen in the star-forming region of the nearby, 3\% Solar metallicity galaxy Leo P \cite{skillman13, mcquinn15} with the MIRI-MRS instrument onboard JWST. 
These observations place a lower limit on the molecular gas content of Leo~P and, combined with our upper limit on carbon monoxide emission from a deep search of this galaxy, demonstrate that MIRI-MRS is sensitive to much smaller molecular gas masses at extremely low metallicity compared to the traditional observational tracer. 
This discovery pushes the maximum metallicity at which purely atomic gas may fuel star formation a factor of two lower, providing crucial empirical guidance for models of star formation in the early universe.}
\bigskip

Observations in the local universe show that star formation occurs in molecular gas in metal-rich galaxies \citep{bolatto11, leroy13}. 
While molecular hydrogen (\molh{}) is the most abundant molecule in these clouds, its lack of an electric dipole moment and low moment of inertia mean that it does not produce observable emission at low temperatures ($\lesssim 100$\,K).
Emission from carbon monoxide (CO) is therefore widely used to trace star-forming, cold molecular gas clouds, but the correlation between \molh{} mass and CO luminosity breaks down at low metallicities. 
The changing balance between CO formation and destruction, primarily driven by the lower dust content of metal-poor gas, combined with the self-shielding of \molh{} results in a smaller CO-emitting region for a molecular cloud of fixed \molh{} mass and consequently a lower observed CO intensity \citep{maloney88, wolfire10, bolatto13}.
The non-detection of CO in star-forming galaxies below 7\% of the Solar value \citep[e.g.,][]{warren15} therefore does not elucidate whether stars can form directly from cold atomic hydrogen (\hi{}) instead of \molh{} in metal-poor environments, as predicted by some models \citep{glover12, krumholz12}.

The temperature of \molh{} in star-forming regions ranges from $<10$\,K in the cold, dense cores of molecular clouds up to several hundred K in their outer layers, where the gas is heated by ultraviolet radiation from short-lived massive stars.
Because rotational \molh{} lines can be collisionally excited in these photodissociation regions (PDRs), it is possible to observe emission from the warm phase of \molh{}, which constitutes $\sim$10--15\% of the total \molh{} mass in typical star-forming galaxies across a wide range of metallicities \citep{roussel07, togi16}.
Such mid-infrared \molh{} emission lines have been detected in moderately metal-poor dwarf galaxies with the Spitzer Space Telescope \citep[e.g.,][]{hunt10, naslim15}, but observations of warm \molh{} remain elusive for star-forming galaxies at extremely low metallicity \citep[i.e., below 5\%\,Solar;][]{mcquinn20}.
It is unclear whether this is because the rotationally excited emission lines are below the sensitivity limit of existing observations or due to a true lack of \molh{}.
CO-emitting molecular clouds are very compact ($\lesssim 2$\,pc) at low metallicity \citep{rubio15, Shi2020}, suggesting that any emission from molecular gas could be strongly diluted in low-angular-resolution data.
JWST's powerful combination of wavelength coverage in the mid-infrared, high spatial resolution, and more sensitive detectors compared to earlier mid-infrared telescopes has opened the possibility of detecting \molh{} emission from an extremely low-metallicity galaxy that is close enough to be spatially resolved.

The 3\% Solar metallicity dwarf galaxy Leo~P is among the most metal-poor star-forming galaxies known in the local universe \citep{skillman13}. Despite its low star-formation rate ($4.3\times10^{-5}\,M_\odot\,\mathrm{yr}^{-1}$; Table~\ref{table1}), the presence of just one O-type star (lifetime $\lesssim10$\,Myr) and several other blue, massive stars unambiguously confirms that Leo~P is actively forming stars \citep{evans19}.
The O star, which powers the only \hii{} region in the galaxy, has been studied in detail with multiwavelength observations that show it has a modest effective temperature of 34\,kK, a high rotation rate, and a weak stellar wind \citep{telford21, telford23, telford24}. 
Leo~P is isolated and has been forming stars at a relatively constant rate over its lifetime \citep{mcquinn15, mcquinn24}, in contrast to similarly metal-poor galaxies like I~Zw~18 \citep{aloisi07} and SBS~0335--052 \citep{izotov97}, which are experiencing bursts of star formation (possibly induced by recent mergers) and therefore have complex and intense radiation fields.
Moreover, these metal-poor starbursts are quite distant; Leo~P is over 10 times closer, just 1.6\,Mpc away.
Its proximity and simple star-forming environment present a unique opportunity to observe the star-formation process at the very low metallicities typical of galaxies in the early universe at high spatial resolution. 

Leo~P's gaseous components have been thoroughly studied via \hi{} 21-cm emission \citep{giovanelli13, bernstein-cooper14} and CO observations \citep{warren15}, which failed to detect molecular gas down to a CO luminosity limit of $L_\mathrm{CO} \leq 2900$\,K\,km\,s$^{-1}$\,pc$^2$.
This motivated us to conduct deeper observations of Leo~P with the Atacama Large Millimeter Array (ALMA), which again resulted in a non-detection, but placed an upper limit on the molecular gas content two orders of magnitude more stringent: $L_\mathrm{CO} \leq 43.6$\,K\,km\,s$^{-1}$\,pc$^2$ (see Table~\ref{table1} and Methods).
Yet, because the ratio of molecular gas mass to CO luminosity (\aco{}) is expected to be large at low metallicity \citep[e.g.,][]{leroy11, glover12}, this deep limit on \lco{} still allows for up to $1.2\times10^4$\,\msun{} of \molh{}.
Not even these state-of-the-art ALMA data can confirm or conclusively rule out molecular gas in Leo~P.

\begin{figure}[!t]
  \includegraphics[width=\linewidth]{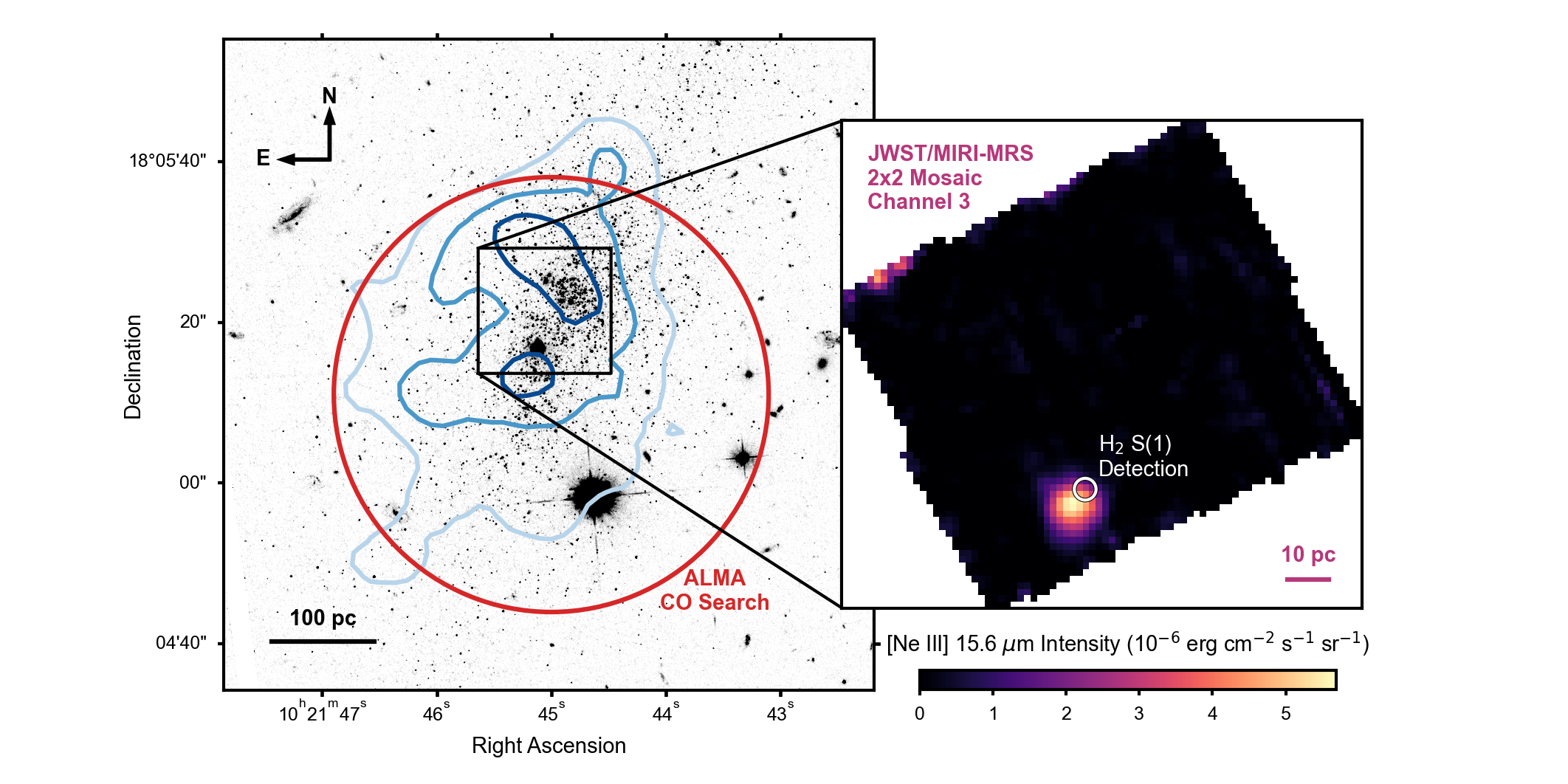}
  \vspace{-14.5pt}
\caption{\textbf{Warm H$\bm{_2}$ emission detected in Leo~P's star-forming region.} A Hubble Space Telescope image of Leo~P in the F475W filter \citep{mcquinn15} is shown in greyscale with contours overlaid to show \hi{} column densities at 8$''$ resolution, where values of 1.6, 2.8, and $4.0\times10^{20}$\,cm$^{-2}$ correspond to increasingly darker shades of blue \citep{bernstein-cooper14}. The bright \hii{} region is visible as the black circular region just north of the southern peak in the \hi{} column density. The red circle indicates the field of view of our ALMA observation that placed a deep upper limit on the CO luminosity (see Table~\ref{table1} and Methods).
The inset image shows a map of the [Ne\,\textsc{iii}]\,$\lambda$15.6\,\mic{} nebular emission line constructed from our JWST MIRI-MRS channel 3 data cube, where the ionized gas in the \hii{} region produces the brightest emission (pink/yellow colors). The white open circle encloses the region where the rotationally excited \molh{} S(1) line at 17.03\,\mic{} is detected in the MIRI-MRS observations (Figure~\ref{fig:leop_h2lines}).
\label{fig:leop_image}}
\end{figure}

We then observed Leo~P with JWST's MIRI-MRS instrument to search for emission from rotationally excited \molh{} (see Methods). 
Figure~\ref{fig:leop_image} shows a Hubble Space Telescope image of Leo~P in greyscale, with blue contours showing the distribution of \hi{} from Very Large Array 21-cm observations  \citep{bernstein-cooper14}.
The region that we searched with ALMA for CO emission is shown as the red circle.
The inset image shows a map of the [Ne\,\textsc{iii}]\,$\lambda$15.6\,\mic{} nebular line intensity constructed from the MIRI-MRS channel 3 data cube, where a spectrum spanning 11.55--17.98\,\mic{} is measured in each spaxel (or spatial pixel).
This line is among the brightest in the MIRI-MRS data, and \hii{} region is visible as the highest-intensity (pink/yellow) spaxels.

We searched the channel 3 data for the \molh{} S(1) emission line at 17.03\,\mic{}, which is expected to be the most easily detectable of the mid-infrared \molh{} rotational transitions (see Methods). 
This revealed \molh{} S(1) emission detected above the 3-$\sigma$ level across six contiguous spaxels near the \hii{} region, unambiguously confirming the presence of molecular gas in the star-forming region of this extremely metal-poor galaxy.
The total area of those six spaxels (each 0.2$''$ on a side) is slightly smaller than the area of one spatial resolution element at 17\,\mic{} (0.67$''$ full width at half maximum, or FWHM). 
Thus, we use the center of those six spaxels to define the location of the unresolved \molh{} S(1) emission, and adopt a diameter equal to the FWHM at 17\,\mic{}; see the white open circle in the inset image of Figure~\ref{fig:leop_image}. 
At the distance of Leo~P, this corresponds to a warm \molh{} cloud with a maximum equivalent radius of 2.6\,pc, similar to the small sizes of the lowest-metallicity cold molecular gas cores detected in CO observations to date \citep{rubio15, Shi2020}. 
The \molh{} S(1) emission is detected at the northwestern edge of Leo~P's only \hii{} region,  consistent with being produced in a classic PDR: a transition zone where the ionized gas meets an \molh{}-bearing cloud.

\begin{figure}[!t]
  \includegraphics[width=\linewidth]{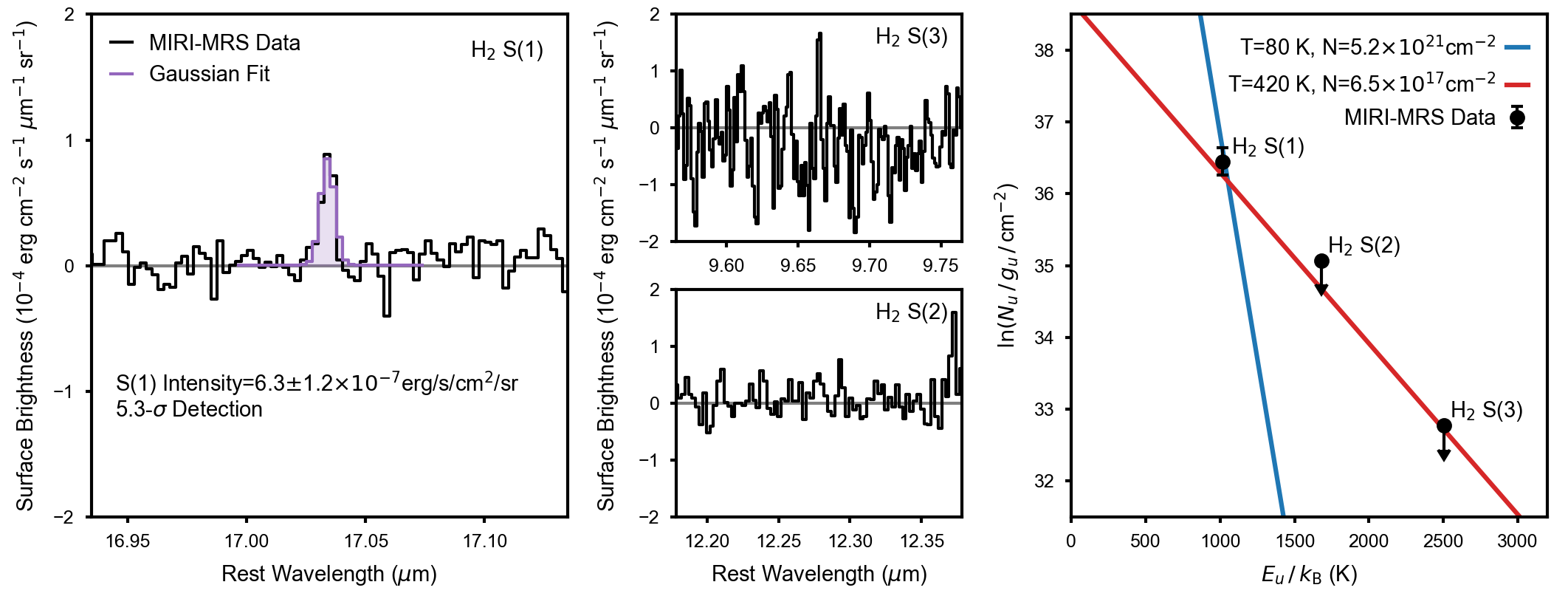}
  \vspace{-14.5pt}
\caption{\textbf{Constraints on the warm \molh{} properties in Leo~P.} \underline{Left}: Shown in black is the average observed MIRI-MRS spectrum within the region indicated by the white circle (0.67$''$ diameter) in the inset image in Figure~\ref{fig:leop_image}, centered on the wavelength of the \molh{} S(1) transition. The best-fit Gaussian model to the data within $\pm700$\,\kms{} of the line center, from which the line intensity is measured, is shown in purple.
\underline{Center}: The average MIRI-MRS spectrum in the same region, now centered on the S(3) and S(2) lines in the top and bottom panels, respectively. Neither line is significantly detected in the observations. 
\underline{Right}: black points show the column density in the upper level of each \molh{} transition ($N_u$), calculated from the measured line intensity or 3-$\sigma$ upper limits (arrows) and normalized by the statistical weight($g_u$), as a function of the  upper level energy expressed as an equivalent temperature ($E_u/k_\mathrm{B}$).
The blue and red lines show the range of single-temperature models consistent with the data (see Methods), with associated temperature ($T$) and total column density ($N$) of warm \molh{} reported in the legend.
The model in blue adopts the lowest temperature at which detectable \molh{} S(1) emission is expected, while the model in red is that with the highest temperature consistent with both the S(1) intensity and S(3) upper limit.
\label{fig:leop_h2lines}}
\end{figure}

The left panel of Figure~\ref{fig:leop_h2lines} presents a portion of the channel 3 MRS spectrum around the \molh{} S(1) line, averaged over the region in which that line is detected (white circle in Figure~\ref{fig:leop_image}).
We measure the line intensity as the area under a Gaussian model (purple) fit to the observed line profile (black), which is securely detected with a signal-to-noise ratio (SNR) of 5.3 (see Methods). 
In addition to the S(1) line, our MIRI-MRS observations cover the wavelengths of three other \molh{} rotational lines arising from excited states of different energies.
The center panels of Figure~\ref{fig:leop_h2lines} show the channel 2 (top) and channel 3 (bottom) MRS spectra in the same spatial region as the S(1) detection centered on the S(3) and S(2) transitions at 9.66\,\mic{} and 12.28\,\mic{}, respectively. 
Neither line is formally detected at $\mathrm{SNR}>3$, but we use the intensity measurement uncertainties to place 3-$\sigma$ upper limits on both lines (see Table~\ref{table1}).
Unfortunately, the noise level is so high at the reddest wavelengths ($\gtrsim24$\,\mic{} in channel 4) that we cannot obtain a useful limit on the S(0)\,28.22\,\mic{} line, which probes the coldest \molh{} that can emit in the mid-infrared.

The measured S(1) intensity of $6.3 \times10^{-7}$\,erg\,s$^{-1}$\,cm$^{-2}$\,sr$^{-1}$ is strikingly low.
To produce such weak emission, either the column density of the emitting warm \molh{} gas must be orders of magnitude lower than typically observed in molecular clouds ($\gtrsim10^{20}$\,cm$^{-2}$) \citep{schinnerer24}, or the temperature of the warm \molh{} component must be very low.
The right panel of Figure~\ref{fig:leop_h2lines} shows how the measured S(1) intensity and S(2) and S(3) upper limits constrain the temperature and total column density of the emitting warm \molh{} (see Methods). 
Black points with error bars (or arrows, for limits) show the column density in the upper energy level of each transition ($N_u$), calculated from the line intensity measurements and normalized by the statistical weight ($g_u$), as a function of the upper level energy ($E_u$) divided by the Boltzmann constant ($k_B$).
The two colored lines show single-temperature models that bound the range of parameters allowed by the data: the blue model has a cold temperature of 80\,K and an associated total column density of $5.2\times10^{21}$\,cm$^{-2}$, while the red model has a higher temperature of 420\,K, and consequently, a total column density nearly four orders of magnitude smaller. If radiative heating by Leo~P's only O star dominates the heating of the observed H$_{2}$, then modeling of the PDR suggests a temperature close to the lower end of this range (see Methods for model details). However, cosmic ray heating or the dissipation of turbulent motions in shocks could potentially yield a higher temperature.

Importantly, because a higher-temperature warm \molh{} cloud with a lower total column density would be inconsistent with the MIRI-MRS data, warm \molh{} with a column density of at least $6.5\times10^{17}$\,cm$^{-2}$ (corresponding to a temperature of 420\,K) must be present in Leo~P.
This implies a minimum warm \molh{} mass of 0.2\,\msun{} within the single spatial resolution element where the S(1) emission line is detected (white circle in Figure~\ref{fig:leop_image}), which has a physical area of 21.7\,pc$^2$. 
While this is quite a small gas mass, it is merely a lower limit; if the emitting warm \molh{} is indeed at a colder temperature of 80\,K, then the required higher column density would translate to a warm \molh{} mass of $1.8\times10^3$\,\msun{}. Moreover, the observational result that just $\sim$10--15\% of the total \molh{} mass is in the warm phase in typical star-forming galaxies \citep{roussel07, togi16} suggests that the minimum total (warm+cold) \molh{} mass in Leo~P is at least 6.5--10 times larger than the lower limit on the warm \molh{} mass, or $\geq1.4$\,\msun{}.
If the warm \molh{} is indeed at 80\,K (i.e., at the lowest end of the temperature range expected to produce S(1) emission), then its higher column density would imply a total \molh{} mass of at least $1.1\times10^{4}$\,\msun{}, which is at the high end of the range of virial masses ($\sim$10$^3$--10$^{4}$\,\msun{}) reported for CO-detected molecular clouds in the metal-poor galaxies WLM and Sextans~B \citep{rubio15, Shi2020}.
Even this higher \molh{} mass implied by the observed rotational line emission is fully consistent with the upper limit on the \molh{} mass of $1.2\times10^{4}$\,\msun{} from our deep ALMA observations (see Methods), but if the emitting gas is at a higher temperature, the \molh{} mass in Leo~P may be orders of magnitude smaller.
The relatively large upper limit is due to the combination of the larger ALMA beam size, which may dilute emission from compact CO-emitting clouds \citep{rubio15, Shi2020}, with the high values of \aco{} expected at low metallicity \citep{leroy11, glover12}.
The much lower \molh{} mass to which our JWST observations are sensitive emphasizes the power of MIRI-MRS as a new tool to detect \molh{} in extremely metal-poor galaxies.

Confirmation of \molh{} in the star-forming region of Leo~P disfavors models in which stars can form in purely atomic gas at 3\% Solar metallicity.
This result provides a valuable constraint on the physics of star formation in the early universe, where the population of surprisingly bright galaxies recently revealed by JWST have posed a major challenge to existing models of galaxy assembly \citep[e.g.,][]{labbe23, boylan-kolchin23, carniani24}.
Our observations in Leo~P show empirically that \molh{} does form even at the extremely low metallicities ($\leq5$\% Solar) observed in many early galaxies \citep[e.g.,][]{topping22, atek24}, an important benchmark for accurately modeling the cooling and chemistry of star-forming gas in that regime.
These findings also support the use of models that tie star formation to \molh{} in galaxy formation simulations \citep[e.g.,][]{christensen14, polzin24} during the early stages of assembly, and for low-metallicity dwarf galaxies at all cosmic epochs.

\begin{table}[!h]
\caption{Observed and Derived Quantities for Leo~P}\label{table1}%
\begin{tabular}{@{}lc@{}}
\toprule
\multicolumn{2}{c}{Galaxy Properties} \\
\midrule
\hii{} Region Right Ascension (J2000) & 10:21:45.1217 \\
\hii{} Region Declination (J2000) & +18:05:16.93 \\
Distance $D$ (Mpc) & $1.62 \pm 0.15$ \\ 
Physical size of 1\,arcsec at $D$ (pc) & $7.85\pm0.73$ \\
Systemic Velocity $v_\mathrm{sys}$ (km\,s$^{-1}$) & $260.8 \pm 2.5$ \\ 
Metallicity $12+\log(\mathrm{O/H})$ & $7.17\pm 0.04$ \\
Stellar Mass \mstar{} (\msun{}) & $2.7\pm0.4\times10^5$ \\  
\hi{} Mass \mhi{} (\msun{}) & $8.1\times10^5$ \\
Star Formation Rate SFR$_{\mathrm{H}\alpha}$ (\msun{}\,yr$^{-1}$) & $4.3\times10^{-5}$ \\
\midrule
\multicolumn{2}{c}{ALMA Observations (Project 2013.1.00397.S)} \\ 
\midrule
Field of View Right Ascension (J2000) & 10:21:45.0000 \\
Field of View Declination (J2000) & +18:05:11.00 \\
Primary Beam FWHM (arcsec) & 54.13 \\
Restoring Beam FWHM (arcsec) & $3.02\times 2.10$ \\
Velocity Resolution (km\,s$^{-1}$) & 0.25 \\
$L_\mathrm{CO}$ (K\,km\,s$^{-1}$\,pc$^2$) & $\leq 43.6$ \\
$M_{\mathrm{H}_2}^\mathrm{cold}$ assuming $\alpha_\mathrm{CO}=285$ (\msun) & $\leq 1.2\times10^4$ \\
\midrule
\multicolumn{2}{c}{MIRI-MRS Observations (JWST-GO-3449)} \\ 
\midrule
\molh{} S(1) Detection Right Ascension (J2000) & 10:21:45.1076 \\
\molh{} S(1) Detection Declination (J2000) & +18:05:17.46 \\
Resolution Element FWHM at 17\,$\mu$m (arcsec) & 0.67 \\
\molh{} S(1) Intensity (erg\,s$^{-1}$\,cm$^{-2}$\,sr$^{-1}$) & $6.3  \pm 1.2\times10^{-7}$ \\
\molh{} S(2) Intensity (erg\,s$^{-1}$\,cm$^{-2}$\,sr$^{-1}$) & $\leq 5.4\times10^{-7}$ \\
\molh{} S(3) Intensity (erg\,s$^{-1}$\,cm$^{-2}$\,sr$^{-1}$) & $\leq 9.1\times10^{-7}$ \\
Warm \molh{} Temperature $T$ (K) & $\leq 420$ \\
Warm \molh{} Column Density $N$ (cm$^{-2}$) & $\geq 6.5\times10^{17}$ \\
Warm \molh{} Mass \mwarm{} (\msun) & $\geq 0.2$ \\
Warm \molh{} Mass if $T=80$\,K  (\msun) &$1.8\times10^3$\\
\botrule
\end{tabular}
\footnotetext{References for galaxy properties: \cite{telford21, mcquinn15, bernstein-cooper14, skillman13, giovanelli13, rhode13, mcquinn24}. Right ascension is reported in hours, minutes, seconds, and declination in degrees, arcminutes, arcseconds. $\alpha_\mathrm{CO}=285$ is a conservatively large predicted value at 3\% Solar metallicity \citep{glover12} (see Methods). }
\end{table}

\section*{Methods}

\subsubsection*{JWST MIRI-MRS Observations and Data Reduction}

We designed JWST program GO-3449 (PI: O.\,G.\,Telford) to detect, or place stringent upper limits on, rotationally excited transitions of warm \molh{} in Leo~P. 
These mid-infrared lines at 9.6--28.2\,\mic{} (see Table~\ref{tab:h2lines}) fall within the wavelength coverage of the the Mid-Infrared Instrument (MIRI)\citep{rieke15, wright23}.
We used MIRI's Medium Resolution Spectroscopy (MRS) observing mode \citep{wells15, argyriou23} to achieve the spectral resolution necessary for faint emission-line measurements.
MIRI-MRS observations are split across 4 integral field units, or channels, where the covered wavelengths increase, the field of view gets larger, and the spatial resolution and sampling get coarser from channels 1 to 4 (see the MIRI-MRS documentation\footnote{\url{https://jwst-docs.stsci.edu/jwst-mid-infrared-instrument/miri-observing-modes/miri-medium-resolution-spectroscopy}} for details).
In each channel, the wavelength coverage is divided into 3 bands (A/SHORT, B/MEDIUM, and C/LONG), and a given band within all 4 channels is observed simultaneously. 
The \molh{} lines of interest span bands A, B, and C, so we observed Leo~P in all 3 bands to cover the full wavelength range accessible to MIRI-MRS. 

Our aim was to search for \molh{} emission across all regions in Leo~P where the emission was likely to be seen, particularly where young stars and the \hii{} region are found
and near the highest observed column densities of \hi{} (see Figure~\ref{fig:leop_image}). 
Because the S(1) line at 17.03\,\mic{} is expected to be detectable across the widest range of possible warm \molh{} temperatures, we selected channel 3 as the primary for mosaic construction (with 10\% overlap between adjacent pointings). 
Good coverage of Leo~P required 4 pointings to construct a $2\times2$ mosaic, given the size of the channel 3 field of view.
We used the SLOWR1 readout pattern, adopted the 4-pt dither pattern optimized for extended sources to improve PSF sampling, and obtained a background observation for each band.

To determine the exposure time necessary to place a 5-$\sigma$ limit on the \molh{} S(1) line, we calculated the expected intensity for a total (warm+cold) \molh{} column density of $10^{21}$\,cm$^{-2}$ with a 10\% warm fraction and temperature of 120\,K, typical of the temperatures and fractions found by the SINGS survey \citep{roussel07}. 
These assumptions gave a predicted intensity of 6.6$\times10^{-7}$\,erg\,s$^{-1}$\,cm$^{-2}$\,sr$^{-1}$.
Using this limiting S(1) intensity in the JWST exposure time calculator\footnote{\url{https://jwst.etc.stsci.edu}} (ETC, v.2), and adopting 12 groups per integration to stay below the 300s limit recommended to mitigate the impacts of cosmic rays, we found that 2 integrations (times the 4 exposures in the dither pattern, for 8 total integrations) were required to reach $\mathrm{S/N}=5$.
In total, this resulted in a total exposure time of 2389\,s for MIRI-MRS band C at each pointing: 4 in the mosaic to cover Leo~P, plus 1 background observation. 
Because the predicted intensities of the other lines are more sensitive to the warm \molh{} temperature, we adopted the same exposure time across all 3 bands for consistency and expected to place weaker limits on the S(0), S(2), and S(3) emission.

The MIRI-MRS observations of Leo~P executed on May 23, 2024.  We retrieved the \texttt{uncal} images from the Mikulski Archive for Space Telescopes (MAST) and processed them using version 1.15.1 of the JWST calibration pipeline with the CRDS context \texttt{jwst1281.pmap} \citep{bushouse2024}. Our reduction followed the steps in the MIRI-MRS batch processing notebook by D.\ Law, retrieved from GitHub in July 2024\footnote{\url{https://github.com/STScI-MIRI/MRS-ExampleNB/blob/main/Flight_Notebook1/MRS_FlightNB1.ipynb}}.  We used a pixel-based background subtraction and processed the Leo~P observations through the \texttt{Detector1}, \texttt{Spec2}, and \texttt{Spec3} stages of the pipeline. We built the cubes in ``ifualign'' coordinates, to later allow for additional row-based background corrections.  Similar to the results presented in \citep{spilker2023}, we found residual stripe artifacts in our reduced MIRI-MRS observations in all four channels. These may be related to imperfect cosmic ray shower removal or to striping and/or detector noise \citep{argyriou23}.  To remove these, again similar to \citep{spilker2023}, we selected $\sim$5 signal-free columns of the cube and found a row-based average background level in each wavelength slice of the cube.  This additional background removal resulted in improved noise characteristics.  We note that the details of this background subtraction are not critical to the detection of narrow emission lines---we detect H$_2$ S(1) at similar S/N with or without the additional row-based background subtraction. It does, however, improve the noise characteristics of the cube and enable more stringent limits on the non-detected H$_2$ lines. 

\begin{table}[t]
\caption{H$_2$ Rotational Lines}\label{tab:h2lines}
\begin{tabular}{cccccccc}
\toprule
\molh{} Line & Transition & Wavelength & MIRI-MRS & MIRI-MRS & $E_u/k_B$& $g_u$ & A  \\
Name & ($\nu=0$)  & (\mic{}) & Channel & Band & (K) & & ($10^{-11}$\,s$^{-1}$) \\
\midrule
S(0) & $J$(2--0) & 28.22 & 4 & C & 510 & 5 & 2.94 \\
S(1) & $J$(3--1) & 17.03 & 3 & C & 1015 & 21 & 47.6 \\
S(2) & $J$(4--2) & 12.28 & 3 & A & 1682 & 9 & 276  \\
S(3) & $J$(5--3) & 9.66  & 2 & B & 2504 & 33 &  984 \\
\botrule
\end{tabular}
\footnotetext{From left to right, the columns report the following parameters used in our observational design and modeling: the short name of each transition; the upper and lower rotational quantum numbers ($J$); the rest wavelength; the MIRI-MRS channel in which the line falls; the MIRI-MRS band; the energy of the upper level expressed as an equivalent temperature; the statistical weight of the upper level; and the Einstein coefficient. Values compiled from \cite{roueff19}.}
\end{table}

\subsubsection*{Emission-Line Flux Measurements}

Of the four pure rotational \molh{} transitions (with no change in the vibrational quantum number $\nu=0$; see Table~\ref{tab:h2lines}) covered by our MIRI-MRS observations, the S(1) transition is expected to be detectable across the widest range of possible temperatures.
Thus, we began by searching individual spaxels in the channel 3 data cube for S(1) emission at 17.03\,\mic{}.
We modeled the wavelengths within $\pm700$\,\kms{} (corresponding to $\pm 0.04$\,\mic{}) of the expected observed wavelength of this transition, accounting for Leo~P's systemic velocity of 260.8\,\kms{}, as a constant continuum level plus a Gaussian emission line.
MIRI-MRS's spectral resolution ($R\sim2400$--3400 over the wavelengths of the \molh{} rotational lines, corresponding to a velocity FWHM of $\sim$90--120\,\kms{}) is insufficient to resolve the expected narrow line width of molecular gas.
Indeed, even emission lines produced in Leo~P's \hii{} region, which are more thermally broadened than the warm \molh{} lines we study here, were unresolved in higher-resolution ($R\sim4000$) Keck Cosmic Web Imager data \citep{telford23}.
Thus, we required the FWHM of the Gaussian emission-line model to be within 1\% of the instrumental broadening, calculated at the wavelength ($\lambda$) of the S(1) line as $R=4603-128 (\lambda / \mu\mathrm{m})$ \citep{argyriou23}.
The root-mean-square (RMS) error in the continuum level surrounding each line was adopted as a conservative estimate of the uncertainties on the observed surface brightness in our fitting with the \texttt{scipy.optimize.least\_squares} routine in Python.
We then calculated the S(1) line intensity in each spaxel as the area under the best-fit Gaussian model and propagated the uncertainties on the inferred parameters to determine the signal-to-noise ratio (S/N) of the measurement. 
After iterating over every spaxel in the channel 3 data cube across all 4 tiles in the mosaic, we found 6 contiguous spaxels with S(1) emission at S/N $>3$. 
We then repeated this procedure to check each spaxel in the channel 3 and channel 2 data cubes for emission in the S(2) and S(3) lines, respectively, but found no significant detections.
We also checked the channel 4 cube for S(0) emission, but found that the noise level was so high at wavelengths $\gtrsim 24$\,\mic{} that any detection of, or useful limits on, the S(0) intensity would be impossible.

The spatial sampling of our reduced MIRI-MRS data cubes is finer than the spatial resolution of the instrument. 
We calculated the FWHM ($\theta$, in arcseconds) of one resolution element at 17.03\,\mic{} as $\theta = 0.033 (\lambda / \mu\mathrm{m}) + 0.106$ \citep{law23}.
The resulting $\theta=0.67''$ corresponds to circular aperture covering the area of 8.8 spaxels (each 0.2$''$ on a side in channel 3).
Thus, the region where S(1) emission is detected is unresolved.
To measure the average S(1) intensity, we took the average spectrum within the single spatial resolution element centered on the middle of the 6-spaxel region in which S(1) is significantly detected (i.e., within the white circle in the right panel of Figure~\ref{fig:leop_image}), then fit a Gaussian plus continuum model to this higher-S/N average spectrum (see the left panel of Figure~\ref{fig:leop_h2lines}).
Again, using the continuum RMS as the measurement uncertainty and standard error propagation techniques, we found a higher-significance S(1) detection at S/N of 5.3.
We repeated that procedure to measure the S(2) and S(3) emission within the same region, but found that $\mathrm{S/N}<3$ for both lines in the averaged spectra.
Thus, we placed 3-$\sigma$ upper limits on those lines equal to 3 times the uncertainties we calculated on their best-fit intensities.

The S(3) transition falls in the shorter-wavelength channel 2, which has smaller spaxels (0.17$''$ on a side) and higher spatial resolution ($\theta=0.42''$ at 9.66\,\mic{}) than the channel 3 cube that covers both the S(1) and S(2) lines (Table~\ref{tab:h2lines}).
For consistency, we fixed the area over which we average the channel 2 spectrum to match that used for channel 3, though this is larger than one resolution element at the wavelength of S(3).
We checked that averaging over a smaller area equal to one resolution element at 9.66\,\mic{} does not yield a significant detection of S(3) emission. 

\subsubsection*{Warm \molh{} Temperature and Mass Limits from the Rotational Lines}

These measurements and upper limits on the \molh{} rotational line strengths in Leo~P constrain the temperature ($T$) and mass (\mwarm{}) of the emitting warm \molh{} \citep[e.g.,][]{burton92}.
The intensity ($I$) of a given \molh{} line is proportional to $N_u$, the column density of \molh{} that is in the upper energy state of that transition.
$N_u$ is calculated from the observed line intensity as:
\begin{equation}
N_u = \frac{4\pi I}{A\Delta E} = \frac{4\pi I \lambda}{Ahc},
\end{equation}
where $A$ is the Einstein coefficient (or emission probability; see Table~\ref{tab:h2lines}) and the change in energy between the upper and lower levels of the transition, $\Delta E$, is equal to $hc/\lambda$, where $h$ is the Planck constant and $c$ is the speed of light. The total column density ($N$) of the emitting warm \molh{} across all energy states can be determined from a model of the form \citep[e.g.,][]{roussel07}:
\begin{equation}\label{eq:NT}
N_u = \frac{g_u N e^{-E_u/k_BT}}{Z(T)},
\end{equation}
where $g_u$ is the statistical weight of the upper level (equal to $(2I+1)(2J+1)$, where $J$ is the rotational quantum number and the total nuclear spin $I$ is 0 for even-$J$ transitions and 1 for odd $J$), $E_u$ is the energy of the upper level, $k_B$ is the Boltzmann constant, and $Z(T)$ is the partition function, which can be approximated accurately by \citep{herbst96}:
\begin{equation}
Z(T) = 0.024\,T\,[1 - e^{-6000/T}]^{-1}.
\end{equation}
The right panel of Figure~\ref{fig:leop_h2lines} shows an excitation diagram constructed from our \molh{} rotational line intensity measurements in Leo~P.
The utility of this $\ln( N_u/g_u)$--$E_u/k_B$ space becomes apparent when we rearrange Equation~\ref{eq:NT} and take the natural logarithm of both sides:
\begin{equation}
\ln(N_u/g_u) =  - (1/T)(E_u/k_B)  + \ln\left(N/Z(T)\right),
\end{equation}
so $T$ is equal to --1 divided by the slope of a linear model in the excitation diagram, and then $N$ can be calculated from the intercept given that value of $T$. 
Because we only detected the S(1) line and placed upper limits on S(2) and S(3), we did not attempt to fit a linear model to the data in Figure~\ref{fig:leop_h2lines}.
Instead, we constructed models varying $T$ in steps of 10\,K across the $T = 80$--1000\,K range over which \molh{} is expected to produce detectable rotational emission \citep{burton92, togi16}.
Then for each $T$, we calculated the range of $N$ allowed within the uncertainties on the S(1) intensity measurement and the upper limit on the S(3) intensity, which is more constraining than the S(2) limit.

We found that models with $T>420$\,K are ruled out by the data, as such models that are consistent with the S(3) upper limit would predict a lower S(1) intensity than observed.
This highest-$T$ model also requires the lowest $N$ to be consistent with our measurements (see Table~\ref{table1}), thus placing a lower limit on the warm \molh{} present.
Since this minimum $N$ was measured from the averaged, spatially unresolved emission over one resolution element, we calculated the corresponding minimum \mwarm{} within the $0.67''$ FWHM (with a radius $r$ equal to 2.63\,pc at the distance of Leo~P) as:
\begin{equation}
\label{eq:warmH2}
M_{\mathrm{H}_2}^\mathrm{warm} = N \pi r^2 m_{\mathrm{H}_2},
\end{equation}
where $m_{\mathrm{H}_2}$ is the mass of an \molh{} molecule, yielding a lower limit of \mwarm\,$\geq0.2$\,\msun{}.
But again, lower $T$ and correspondingly higher $N$ are allowed by the data, so the true warm \molh{} content of Leo~P could be much larger. 
Without a limit on the S(0) line intensity, we cannot determine the lowest allowed $T$ of the emitting \molh{}, so adopt 80\,K as a reasonable threshold below which the S(1) intensity is expected to drop off precipitously and become undetectable \citep{burton92, togi16}. 
If the true $T$ of the emitting \molh{} in Leo~P were 80\,K, then the maximum $N$ consistent with our observations would correspond to $M_{\mathrm{H}_2}^\mathrm{warm}=1.8\times10^3$\,\msun{}.
These \molh{} mass estimates are only for the emitting, warm gas; yet, previous observations of star-forming galaxies at higher metallicity have demonstrated that $\sim$10--15\% of the total \molh{} mass is typically in the warm phase \citep{roussel07, togi16}.
Assuming that a similar temperature distribution holds at 3\% Solar metallicity, then we can scale our \mwarm{} calculations to the implied minimum total (warm+cold) \molh{} mass (\mmolh{}) using a total-to-warm \molh{} ratio of 6.5.
This results in \mmolh{} of $\geq 1.4$\,\msun{} for the lower-limit $N$ corresponding to $T=420$\,K, or $1.1\times10^4$\,\msun{} if $T=80$\,K.

Finally, we note that though warm \molh{} emission is often modeled as arising from two components at two different temperatures \citep[e.g., the PDR Toolbox;][]{kaufman06, pound08, pound11, pound23}, we have too few line detections to constrain all of the parameters in that more complex model, so we simply assume that all of the warm \molh{} is at a single $T$.
We also assume a fixed equilibrium ortho-to-para ratio of 3, as we do not have enough rotational line detections to include this as a free parameter in our model.
While it is possible that the true value may be lower if the emitting \molh{} is out of equilibrium or at a low temperature $\lesssim200$\,K \citep{burton92}, this assumption does not impact our conclusions. 
If the true ortho-to-para ratio is $< 3$, then our measured $N_u/g_u$ for the odd-$J$ (ortho) transitions S(1) and S(3) would be underestimated by the same vertical offset in Figure~\ref{fig:leop_h2lines} \citep[e.g.,][]{sheffer11}.
Because our temperature constraint comes from the range of slopes allowed by the S(1) and S(3) data, it is unchanged by the implied change in normalization of the model.
Our $N_\mathrm{tot}$ lower limits would change in the sense that the corrected values would be larger than what we report here, so the minimum warm and total \mmolh{} that we find in Leo~P are conservatively small.
Moreover, this is at most a factor of three effect because even at 80\,K, the ortho-to-para ratio is $\sim$1.1 \citep{burton92}. 

\subsubsection*{Expected Warm \molh{} Temperature from the O Star's Ultraviolet Radiation}

Observations of star-forming galaxies indicate that most of their \molh{} rotational line emission is excited by massive stars illuminating molecular gas \citep[e.g.,][]{roussel07}.
Because Leo~P contains only one O star at the center of its \hii{} region, the ultraviolet radiation field illuminating the PDR where S(1) emission is detected is unusually well-constrained.
We therefore used the observed properties of the O star to investigate what range of temperatures we should expect for the H$_{2}$ at the location of our detection.

We performed some exploratory PDR modeling using the simple hydrogen-carbon-oxygen chemical model described in Appendix A of \cite{hunter2023}, which is a modified version of a model originally introduced by \cite{gong2017}. In these initial calculations, we adopted a slab geometry and constant density ($n$) for the gas, and computed the temperature and chemical composition as a function of depth into the slab. To fix the strength of the illuminating radiation field, we adopted a separation of 4.4~pc between the O star and the slab, as suggested by the observed angular separation of $0.57''$. The true separation could be larger if the H$_{2}$-emitting gas lies in the foreground or background compared to the O star, but is unlikely to be significantly smaller. Using a model of the ultraviolet spectrum that was fit to the O star's far-ultraviolet through near-infrared spectral energy distribution observed with the Hubble Space Telescope \citep{telford21}, we computed an incident photon flux of $F \simeq 1.8 \times 10^{9} \: {\rm cm^{-2}} \: {\rm s^{-1}}$ in the 11.2--13.6~eV energy band at the position of the slab, corresponding to a radiation field strength of approximately $\chi \simeq 150$ in Habing units \citep{habing1968}. For the purposes of our modeling, we conservatively adopted a slightly larger field strength of $\chi = 200$ to account for the uncertainty in the exact location of the front of the slab. 

We fixed the metallicity of the gas to 3\% of the Solar value and assumed for simplicity that the elemental abundance ratio of C/O is the same as that in the local warm neutral medium \citep{sembach2000}. We investigated the effect of adopting a smaller C/O ratio, and found that a factor of 2 decrease leads to a 25--30\% higher temperature at the lowest densities, but negligible differences above $n \sim 1000 \: {\rm cm^{-3}}$. We also assumed a dust-to-gas ratio of 3\% of the local value and adopted a treatment of photoelectric heating that assumes the presence of polycyclic aromatic hydrocarbons (PAHs) \citep{bakes1994,wolfire2003}. In reality, it is plausible that the actual dust-to-gas ratio in Leo~P is smaller than the value one gets by scaling linearly with metallicity \citep{remy-ruyer14} and it is also likely that photoheating from PAHs is negligible due to the sharp decrease in PAH abundance with metallicity \citep{draine07, engelbracht08, whitcomb24}; both effects would reduce the photoelectric heating rate and hence reduce the gas temperature. 

\begin{figure}
\begin{center}
  \includegraphics[width=\linewidth]{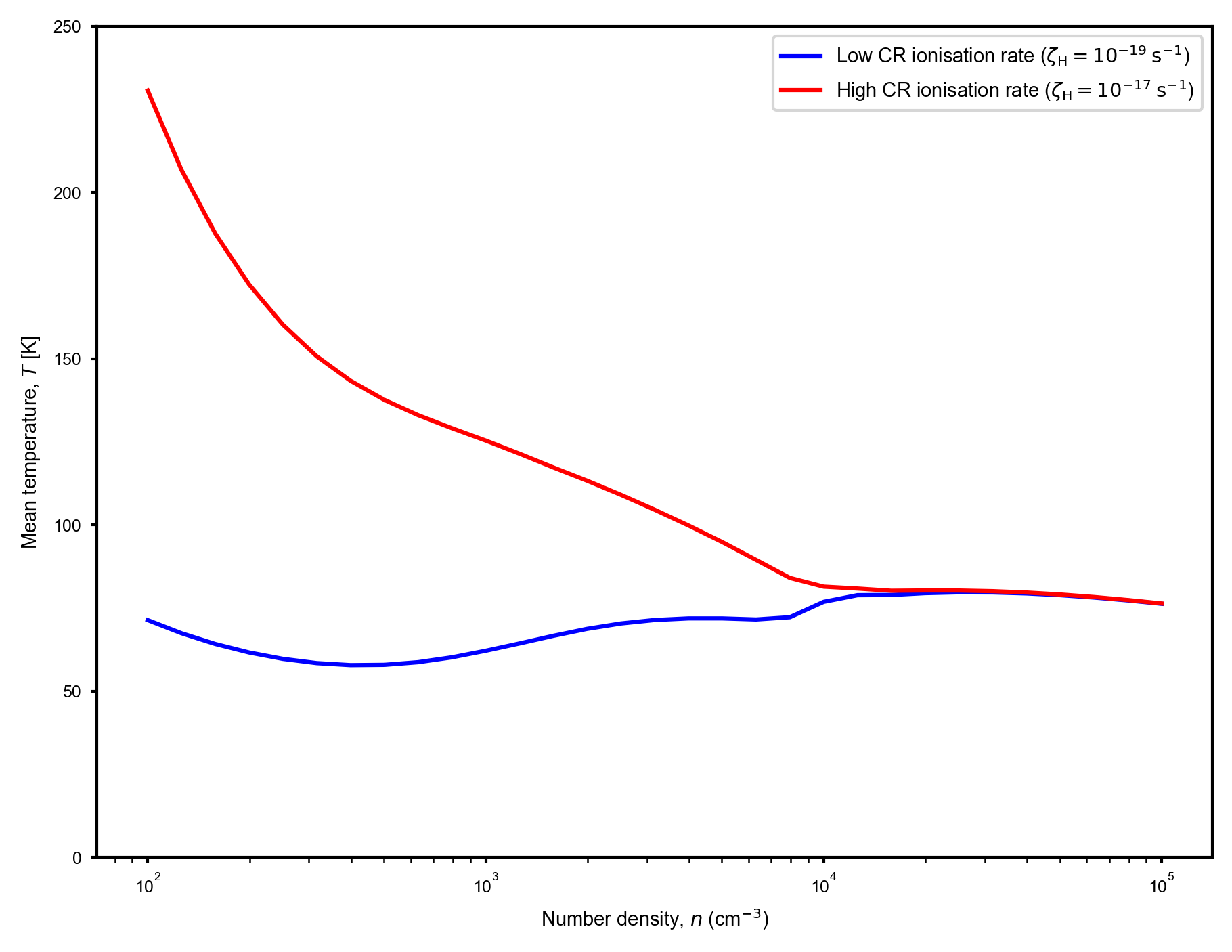}
\end{center}
\caption{H$_2$ S(1) emission-weighted mean temperature as a function of density for our simple one-dimensional PDR models of the warm H$_{2}$ in Leo P (see text for details). These calculations assume a slab symmetry for the emitting gas, with a thickness $L = 1$~pc and a uniform density as indicated in the figure. The slab is illuminated by the radiation field produced by the O star in Leo~P, which has a strength of $\chi = 200$ in Habing units \citep{habing1968}. Results are shown for two different values for the cosmic ray (CR) ionisation rate: $\zeta_{\rm H} = 10^{-19} \: {\rm s^{-1}}$ (blue line) and $\zeta_{\rm H} = 10^{-17} \: {\rm s^{-1}}$ (red line). 
\label{fig:H2temp_model}
}
\end{figure}

In Figure~\ref{fig:H2temp_model}, we show the average temperature ($T$) of the slab as a function of $n$. These calculations assume a slab of thickness $L = 1$~pc, but the results are insensitive to variations of a factor of 2 in this value. The average $T$ shown here is weighted by H$_{2}$ S(1) emission, estimated assuming LTE level populations for the low-lying rotational levels of H$_{2}$. Weighting instead by H$_{2}$ mass yields similar results at $n < 10^{4} \: {\rm cm^{-3}}$, but lower temperatures at higher densities where the mass-weighted average is primarily sensitive to the temperature of the cold H$_2$. We show results for two sets of calculations: one in which we fixed the cosmic ray ionisation rate of atomic hydrogen to a low value, $\zeta_{\rm H} = 10^{-19} \: {\rm s^{-1}}$ (blue line) and a second in which we adopted a much larger value of  $\zeta_{\rm H} = 10^{-17} \: {\rm s^{-1}}$ (red line). Our choice of the low value here is designed to be representative of conditions in which we expect photoelectric heating to dominate, but our results are insensitive to the precise value chosen provided that $\zeta_{\rm H} < 10^{-18} \: {\rm s^{-1}}$. Our choice of the high value is motivated by recent numerical simulations of star formation in a metal-poor system with a star formation rate two orders of magnitude higher than Leo~P \citep{hu2023}. 
We see that heating by the photoelectric effect plus a low level of cosmic rays yields an average temperature that is less than 100\,K at all densities considered here, suggesting that our S(1) detection is consistent with being produced in gas at the low end of the temperature range expected to excite \molh{} rotational lines, given the properties of the O star.
Significantly increasing the cosmic ray ionisation rate does produce gas with $T > 100$\,K at densities below a few 1000\,${\rm cm^{-3}}$, but even here the temperature is towards the bottom end of the range allowed by the S(1) detection and S(3) non-detection. 
We consider such a high value of $\zeta_{\rm H}$ to be unlikely, given the low star formation rate of Leo~P \citep{girichidis2020}, but in the absence of a direct observational constraint, we cannot exclude this value. We also note that heating by an incident X-ray flux or by turbulent dissipation in shocks (if modeled as a uniform heating term) has a very similar effect. 

Though we lack observational constraints on $\zeta_{\rm H}$ or the turbulent dissipation rate, it is unlikely that cosmic rays, X-rays, or shocks contribute significantly to the PDR heating given the apparently simple environment in Leo~P.
Moreover, our modeling shows that even an implausibly high CR ionisation rate cannot produce temperatures at the high end of the range allowed by our observations of the H$_{2}$ rotational lines. 
Because the O star is expected to dominate the PDR heating, our modeling suggests that the temperature of the warm H$_{2}$ in Leo~P is likely to be towards the lower end of the allowed range.

\subsubsection*{ALMA Observations and Upper Limits on $L_\mathrm{CO}$ and \mmolh{}}

The ALMA observations of Leo P were taken on December 7th, 2014 (Project 2013.1.00397.S; PI: S.\,Warren) targeting the 115 GHz $^{12}$CO ($J$=1--0) line in band 3 with the 12\,m array and an integration time of 10705\,s or 2.974\, hours. The calibrators were J1058+0133 for the bandpass calibrator, Titan for the flux calibrator, and J1012+2312 for the phase calibrator. The observations consisted of a single pointing that covered the peak of the \hi\ distribution, where the field of view of the FWHM of the primary beam is shown as the red circle in Figure \ref{fig:leop_image}. We reimaged the raw calibrated data with version 4.2.2 of CASA (Common Astronomy Software Applications \cite{casas2022}) using the \texttt{clean} task and setting the image weight to natural weighting to optimize sensitivity. The restoring beam of the resulting data cube is $\ang{;;3.02}$ by $\ang{;;2.10}$ and the velocity resolution is 0.25\,km\,s$^{-1}$. We find no robust CO(1--0) detections in the data, and note that these ALMA observations are limited by the relatively large beam size of $\ang{;;3.02}$ by $\ang{;;2.10}$, which may dilute the emission from the very small ($<\ang{;;1}$) CO clouds expected from the size of the H$_2$ region (see Figure \ref{fig:leop_image}). Still, we can use measurements of the noise to estimate a robust upper limit on the molecular gas mass in Leo~P for comparison to the JWST MIRI-MRS detection of \molh{} S(1) emission. 

We took two approaches to estimate the upper limit on the CO luminosity (\lco{}). The first approach follows a similar procedure to \cite{warren15} and \cite{Leroy2007}, enabling a direct comparison to the previous \lco{} limit reported for Leo~P. We smoothed the data cube to a velocity resolution of 2.5\,km\,s$^{-1}$ to match the average CO linewidth of molecular clouds observed in the low-metallicity galaxies WLM and Sextans B \citep{rubio15, Shi2020}, as well as the velocity resolution of \hi{} observations of Leo~P \citep{bernstein-cooper14}. We calculated the upper limit on the CO intensity ($S_{\rm CO}$) by measuring the RMS across the velocity channels that contain \hi\ emission (250 -- 270\,km\,s$^{-1}$), multiplied by the velocity channel width ($\Delta v = 2.5\,{\rm km\,s^{-1}}$), then multiplied by a factor of 4 to achieve a 4-$\sigma$ upper limit. The CO luminosity is given by:
\begin{equation}
\label{eq:LCO}
    L_{\rm CO} = 2453 D^2 S_{\rm CO} = 9812 D^2 \Delta v \, \rm{RMS}\ [\,K\,km\,s^{-1}\,pc^2]i
\end{equation}
where the distance $D = 1.62\,$Mpc (Table \ref{table1}). The ALMA observations result in an upper limit of $L_\mathrm{CO}\leq83.4$\,K\,\,km\,s$^{-1}$\,pc$^2$, which is almost two orders of magnitude deeper than the previous upper limit of $L_\mathrm{CO}\leq2900$\,K\,\,km\,s$^{-1}$\,pc$^2$ presented in \cite{warren15} using Combined Array for Millimeter-wave Astronomy (CARMA) observations of Leo~P.

The second approach to calculate the \lco{} upper limit uses a stacking procedure outlined in \cite{Schruba2011, Schruba2012} that assumes the velocity of any undetected CO clouds will be similar to the peak velocity of the \hi\ emission, as is seen in other dwarf galaxies \citep[e.g.,][]{rubio15, Schruba2012}. We first extracted a 100$\times$100\,pixel ($\ang{;;70}\times\ang{;;70}$) square region centered on the \hii\ region and the \molh{} S(1) detection. We regridded and interpolated the high-resolution $\ang{;;8}$ \hi\ data cube presented in \cite{bernstein-cooper14} to match the spectral resolution and pixel scale of the ALMA CO data cube. With the CO and \hi\ data on the same frame, we fit the \hi\ spectrum in a $10\times10$ pixel sub-region with a Gaussian model to identify the peak velocity. Each CO spectrum in a given sub-region was then shifted to this peak \hi\ velocity and the process was repeated over the full 100-pixel-square region. 
We averaged the sub-region spectra to produce a final stacked spectrum, which decreased the noise by about a factor of 2. 
Using Equation~\ref{eq:LCO}, we calculated the 4-$\sigma$ \lco{} limit by measuring the RMS within the \hi\ emission velocity bounds (250--270\,km\,s$^{-1}$) and the velocity resolution of the CO data ($\Delta v = 0.25 \, {\rm km}\,s^{-1}$) used in stacking. The final upper limit from stacking is $L_{\rm CO} \leq 43.6\,{\rm K\,\,km\,s^{-1}\,pc^2}$, a factor of 2 lower than the upper limit calculated above. 
We adopt this upper limit from stacking in our calculations below as it is more sensitive than the traditional upper limit, which we report to facilitate comparison to the results in \cite{warren15}. 

The observed $L_{\rm CO}$ is commonly used to determine the molecular hydrogen mass ($M_{\mathrm{H}_2}$), but requires an assumption of a CO-to-$\rm H_{2}$ conversion factor, $\alpha_{\rm CO}$:
\begin{equation}
\label{eq:alphaCO}
    M_{\mathrm{H}_2} = \alpha_{\rm CO} L_{\rm CO}.
\end{equation}
While $\alpha_{\rm CO} = 4.3\,M_{\odot}\, \rm (K\,km\,s^{-1}\,pc^2)^{-1}$ at Solar metallicity, the CO-to-$\rm H_{2}$ conversion factor is known to vary substantially with metallicity \cite[e.g.,][]{bolatto13}. Galaxies with metallicities below Solar tend to have large reservoirs of ``CO-dark'' molecular gas, defined as areas of the interstellar medium where H$_2$ is present but the CO molecule is dissociated \cite[e.g.,][]{Grenier2005, wolfire10, glover12, Jameson2018, Madden2020}. It is therefore difficult to place a stringent constraint on the total molecular gas mass in an extremely low-metallicity galaxy like Leo~P, particularly because it remains unclear how \aco{} varies as a function of metallicity. To place a conservatively large upper limit on \mmolh{}, we use a theoretical CO-to-H$_2$ conversion factor from \cite{glover12}, whose 3\% Solar metallicity models predict $\alpha_{\rm CO} =  285\,M_{\odot}\, \rm (K\,km\,s^{-1}\,pc^2)^{-1}$. This results in an upper limit of $M_{\mathrm{H}_2} \leq 1.2 \times 10^4\, M_{\odot}$ in Leo~P, using the most sensitive \lco{} limit from stacking. 
For comparison, adopting instead the $\alpha_\mathrm{CO}=70$ measured at 20\% Solar metallicity in the Small Magellanic Cloud \citep{leroy11} would imply $M_{\mathrm{H}_2} \leq 3.1\times10^3$\,\msun{}, but \aco{} is expected to be higher than this at the much lower 3\% Solar metallicity of Leo~P.
Combining Leo~P's \hi{} mass (Table~\ref{table1}) with the maximum \mmolh{} of $1.2 \times 10^4\, M_{\odot}$ yields an upper limit on the molecular-to-atomic gas ratio of $\leq 0.015$, lower than the range observed in other nearby, star-forming dwarf galaxies \citep{leroy09}.

\bigskip

\backmatter

\bmhead{Acknowledgements}

Based on observations with the NASA/ESA James Webb Space Telescope obtained from MAST at the Space Telescope Science Institute, which is operated by the Association of Universities for Research in Astronomy, Incorporated, under NASA contract NAS5-26555.
This paper makes use of ALMA data.
ALMA is a partnership of ESO (representing its member states), NSF (USA) and NINS (Japan), together with NRC (Canada), MOST and ASIAA (Taiwan), and KASI (Republic of Korea), in cooperation with the Republic of Chile. 
The Joint ALMA Observatory is operated by ESO, AUI/NRAO and NAOJ.
The National Radio Astronomy Observatory is a facility of the National Science Foundation operated under cooperative agreement by Associated Universities, Inc.
This research used NASA Astrophysics Data System Bibliographic Services, adstex, and the arXiv preprint server.
The following software was used in this analysis: Astropy \citep{astropy13, astropy18, astropy22}, iPython \citep{perez07}, Matplotlib \citep{hunter07}, NumPy \citep{van-der-walt11, harris20}, SAOImageDS9 \citep{joye03}, SciPy \citep{virtanen20}, CASA \cite{casas2022}, SpectralCube \cite{speccube}.

\bmhead{Funding}
Support for this work was provided by NASA through grant JWST-GO-3449 from the Space Telescope Science Institute under NASA contract NAS5-26555.
O.\,G.\,T. acknowledges support from a Carnegie-Princeton Fellowship through Princeton University and the Carnegie Observatories.
This research was supported in part by grant NSF PHY-2309135 to the Kavli Institute for Theoretical Physics (KITP). S.\,C.\,O.\,G. acknowledges funding from the European Research Council via the ERC Synergy Grant ``ECOGAL'' (project ID 855130) and from the German Excellence Strategy via the Heidelberg Cluster of Excellence ``STRUCTURES'' (EXC 2181 - 390900948). A.\,D.\,B. acknowledges support from the NSF under award AST-2108140.

\bmhead{Author contributions} O.\,G.\,T., K.\,M.\,S., and K.\,B.\,W.\,M. developed the idea for the JWST proposal.
O.\,G.\,T. (PI of JWST-GO-3449) analyzed the \molh{} emission lines in the MIRI-MRS observations of Leo~P, made Figure~\ref{fig:leop_image} and Figure~\ref{fig:leop_h2lines}, and wrote the manuscript. 
K.\,M.\,S. led the design of the JWST observations and reduced the MIRI-MRS data.
S.\,C.\,O.\,G. modeled the PDR to constrain the emitting \molh{} temperature and made Figure~\ref{fig:H2temp_model}.
E.\,J.\,T. reduced the ALMA data and calculated the upper limit on \lco{} in Leo~P.
All authors contributed to interpretation of the scientific results and to the manuscript. 

\bmhead{Competing interests}
The authors declare no competing interests.

\bmhead{Data availability}
JWST MIRI-MRS data from program GO-3449 will be available to download from the Mikulski Archive for Space Telescopes (\url{https://mast.stsci.edu}) as of May 23, 2025. 
Data for ALMA Project 2013.1.00397.S are available to download from the ALMA Archive (\url{https://almascience.nrao.edu/alma-data}). 
Results from the PDR modeling are available from the authors on request.
All other data in this paper have been previously published.

\bmhead{Code availability}
Code to analyze the data and produce figures in this paper will be made available upon reasonable request to the corresponding author (O.\,G.\,T.; {\color{blue}grace.telford@princeton.edu}).


\end{document}